\documentclass[twocolumn,showpacs,aps,prd,superscriptaddress]{revtex4}

\usepackage{graphicx}
\usepackage{dcolumn}
\usepackage{amsmath}
\usepackage{epsfig}

\input babarsym

\def\pipi     {\ensuremath{\pi\pi}}

\def\akpi  {\ensuremath{{\cal A}_{K\pi}}}

\def\spipi {\ensuremath{S_{\pi\pi}}}
\def\cpipi {\ensuremath{C_{\pi\pi}}}
\def\de {\ensuremath{\Delta E}}

\newcommand{\BABARPubYear}    {06}
\newcommand{\BABARPubNumber}  {047}

\newcommand{\SLACPubNumber} {12030}

\def\figurebox#1#2#3{%
    \def\arg{#3}%
    \ifx\arg\empty
    {\hfill\vbox{\hsize#2\hrule\hbox to #2{\vrule\hfill\vbox to #1{\hsize#2\vfill}\vrule}\hrule}\hfill}%
    \else
    {\hfill\epsfbox{#3}\hfill}%
    \fi}

\begin{document}

\preprint{\babar-PUB-\BABARPubYear/\BABARPubNumber} 
\preprint{SLAC-PUB-\SLACPubNumber} 

\begin{flushleft}
\babar\ Analysis Document \#1391, Version 15 \\
\babar-PUB-\BABARPubYear/\BABARPubNumber \\

SLAC-PUB-\SLACPubNumber \\
\end{flushleft}

\title{
{\large \bf
Improved Measurements of the Branching Fractions for {\boldmath $\Bz\to\pip\pim$} and 
{\boldmath $\Bz\to\Kp\pim$}, and a Search for {\boldmath $\Bz\to\Kp\Km$}} 
}

%
\author{B.~Aubert}
\author{R.~Barate}
\author{M.~Bona}
\author{D.~Boutigny}
\author{F.~Couderc}
\author{Y.~Karyotakis}
\author{J.~P.~Lees}
\author{V.~Poireau}
\author{V.~Tisserand}
\author{A.~Zghiche}
\affiliation{Laboratoire de Physique des Particules, F-74941 Annecy-le-Vieux, France }
% ...
\collaboration{The \babar\ Collaboration}
\noaffiliation

\date{\today}

\begin{abstract}
We present measurements of the branching fractions for the charmless two-body decays 
$\Bz\to\pip\pim$ and $\Bz\to\Kp\pim$, and a search for the decay $\Bz\to\Kp\Km$.  We
include the effects of final-state radiation from the daughter mesons for the first time, 
and quote branching fractions for the inclusive processes $\Bz\to h^+ h^{\prime -} n\gamma$, where 
$h$ and $h^\prime$ are pions or kaons.  The maximum value of the sum of the energies of the 
$n$ undetected photons, $E_\gamma^{\rm max}$, is mode-dependent.
Using a data sample of approximately $227$ million \upsbb\ decays collected 
with the \babar\ detector at the \pep2\ asymmetric-energy \epem\ collider at SLAC, we 
measure:
\begin{eqnarray*}
{\cal B} (\Bz \to \pip\pim\, n\gamma;\: E_{\gamma}^{\rm max}=150\mev) & = & (5.1\pm 0.4\pm 0.2)\times 10^{-6},\\ 
{\cal B} (\Bz \to \Kp\pim\, n\gamma;\: E_{\gamma}^{\rm max}=105\mev) & = & (18.1\pm 0.6\pm 0.6)\times 10^{-6}, \\
{\cal B} (\Bz \to \Kp\Km\, n\gamma;\: E_{\gamma}^{\rm max}=59\mev) & < & 0.5 \times 10^{-6}\,
(90\%~{\rm confidence~level}),
\end{eqnarray*}
where the first uncertainty is statistical and the second is systematic.  
Theoretical calculations can be used to extrapolate from the above measurements the non-radiative 
branching fractions, ${\cal B}^0$.  Using one such calculation, we find:
\begin{eqnarray*}
{\cal B}^0 (\Bz \to \pip\pim) & = & (5.5\pm 0.4\pm 0.3)\times 10^{-6},\\ 
{\cal B}^0 (\Bz \to \Kp\pim) & = & (19.1\pm 0.6\pm 0.6)\times 10^{-6}, \\
{\cal B}^0 (\Bz \to \Kp\Km) & < & 0.5 \times 10^{-6}\,
(90\%~{\rm confidence~level}).
\end{eqnarray*}
Meaningful comparison between theory and experiment, as well as combination of measurements from 
different experiments, can be performed only in terms of these non-radiative quantities.

\end{abstract}


\maketitle

Charmless hadronic two-body $B$ decays to pions and kaons provide a wealth of 
information on \CP\ violation in the $B$ system, including all angles of the 
unitarity triangle.  The time-dependent \CP\ asymmetries in the $\pi\pi$ system can be 
used to estimate the angle $\alpha$~\cite{pipialpha}; the decay rates for the 
$K\pi$ channels provide information on $\gamma$~\cite{gammaKpi}. 
Recently, direct \CP\ violation in decay was 
established in the $B$ system through observation of a significant rate asymmetry between 
$\Bz\to\Kp\pim$ and $\Bzb\to\Km\pip$ decays~\cite{babarAkpi,belleAkpi}.  
As $B$ physics experiments accumulate much larger data sets, charmless two-body $B$ decays 
will continue to play a fundamental role in testing the standard model description of 
\CP\ violation.
Measurements of branching fractions for
all the charmless two-body decays are invaluable in testing the
various theoretical approaches to the underlying hadron dynamics~\cite{thy}.
We present measurements of branching fractions 
for the decays $\Bz\to\pip\pim$ and $\Kp\pim$~\cite{cc}, and a search for the decay $\Bz\to\Kp\Km$
using a data sample about $2.5$ times larger than that used for the most precise, previously 
published measurements~\cite{BaBarsin2alpha2002,belleBR,belleKK} of these quantities.

As radiative corrections have already proved to be important in precise determinations
of interesting quantities in the context of kaon physics \cite{Ambrosino:2005ec}, we account
for them in this analysis as well. We can relate the observable
decay rates $\Gamma_{hh^{'}} (E_\gamma^{\rm max})$ 
for $\Bz \to  h^+ h^{'-} n \gamma$ (and thus the branching fractions) to the theoretical non-radiative
widths $ \Gamma_{hh^{'}}^0 $, using the energy-dependent
correction factors $ G_{hh^{'}}(E_\gamma^{\rm max};\mu) $ \cite{QED}
\begin{eqnarray}
\Gamma_{hh^{'}} (E_\gamma^{\rm max})  & =  &
\Gamma(\Bz \to h^+ h^- n \gamma ) | _
{ \sum E_{\gamma} < E_\gamma^{\rm max}} \cr
 & = & \Gamma_{hh^{'}}^0(\mu) \, G_{hh^{'}} ( E_\gamma^{\rm max} ; \mu ),
\end{eqnarray}
where $E_\gamma^{\rm max}$ is the maximum value allowed for the sum of the undetected photon energies and
$\mu$ is the renormalization scale at which $\Gamma_{hh^{'}}^0$
and $G_{hh^{'}} ( E_\gamma^{\rm max} )$ are calculated (the product being independent of $\mu$).
Extracting $ \Gamma_{hh^{'}}^0 $ allows a more meaningful comparison with theoretical 
calculations and also between different experimental results. 
Additionally,
for $E_\gamma^{\rm max}$ at the kinematic limit,
$G$ approaches unity (to order $ \alpha_{\rm QED} / \pi $),
so that the $ \Gamma_{hh^{'}}^0 $, and the corresponding
branching fractions, can be interpreted theoretically in a cleaner way.

The data sample used for this analysis contains $(226.6\pm 2.5)\times 10^6$ 
$\Y4S\to\BB$ decays collected by the \babar\ detector~\cite{ref:babar} at the 
SLAC PEP-II $\epem$ asymmetric-energy storage ring.  The primary detector 
components used in the analysis are a charged-particle tracking system 
consisting of a five-layer silicon vertex detector and a 40-layer drift chamber 
surrounded by a $1.5$-T solenoidal magnet, an electromagnetic calorimeter 
comprising $6580$ CsI(Tl) crystals, and a dedicated particle-identification 
system consisting of a detector of internally reflected 
Cherenkov light providing at least 3 $\sigma$ $K$--$\pi$ separation over the range of laboratory 
momentum relevant for this study ($1.5$--$4.5\gevc$).

The data sample used in this analysis is similar to that used in the \babar\ measurements 
of direct \CP\ violation in $\Bz\to\Kp\pim$~\cite{babarAkpi} and time-dependent
\CP-violating asymmetry amplitudes $\spipi$ and $\cpipi$ in $\Bz\to\pip\pim$~\cite{BaBarsin2alpha2005} (the
reader is referred to those references for further details of the analysis
technique).  Event selection criteria are identical to those used in the \CP\ analyses
~\cite{babarAkpi,BaBarsin2alpha2005}, except that we remove 
the requirement on the difference in the decay times ($\Delta t$) between the two $B$ 
mesons in order to minimize systematic uncertainties on the branching fraction measurements.  

We identify $\Bz\to h^+h^{\prime -}$ ($h,h^{\prime}= \pi$ or $K$) candidates 
with selection requirements on track and Cherenkov angle ($\theta_c$) quality, $B$ decay 
kinematic variables, and event topology. 
The final sample contains $69264$ events and 
is defined by requirements on two kinematic variables:  (1) the 
difference $\de = E_B^{\ast} - \sqrt{s}/2$  between the reconstructed energy of the $B$ candidate
in the $\epem$ center-of-mass (CM) frame and $\sqrt{s}/2$; and (2) the beam-energy 
substituted mass 
$\mes = \sqrt{(s/2 + {\mathbf {p}}_i\cdot {\mathbf {p_B}})^2/E_i^2- {\mathbf {p_B}}^2}$.
Here, $\sqrt{s}$ is the total CM energy, and the $B$ momentum ${\mathbf {p_B}}$ 
and the four-momentum $(E_i, {\mathbf {p_i}})$ of the $\epem$ initial state are 
defined in the laboratory frame.  
To simplify the analysis, we use the pion mass for all tracks in the track reconstruction and 
the calculation of the kinematic variables.
We select those $B$ candidates with $\left | \de\right | < 150\mev$, and 
$5.20 < \mes < 5.29\gevcc$.

The efficiencies of the selection criteria are determined in samples of 
\geant-4 based~\cite{geant} Monte Carlo (MC) simulated signal decays, where we
include the effects of electromagnetic radiation from 
the final-state charged particles using the {\tt PHOTOS} simulation package~\cite{photos}.

We compare the performance of our simulation with a scalar QED calculation\cite{QED} 
resummed to all orders of $\alpha_{\rm QED}$.  Among events selected by the $\left | \de\right | < 150\mev$ requirement,  
the MC simulation and Ref.~\cite{QED} predict different fractions of events with photons with energy 
below $2.6 \mev$, the soft photon energy cut-off used in our simulation (see Tab.~\ref{tab:first_bin}).   
We therefore reweight the $\de$ distributions for each mode to account 
for this different fraction of radiating events and use these reweighted distributions
in the final maximum likelihood fit.  The difference in event yields obtained with the original distributions and with 
the reweighted ones is used to evaluate the associated systematic error, and it is found to be negligible.
\begin{table}[!tbp]
\begin{center}
\begin{tabular}{lcc} 
\hline\hline
   Mode    & MC &  QED calculation \\
\hline
$\pi^+ \pi^-$ &  $89.8\pm 0.1 $    &$88.8\pm 0.5$ \\
$K^+ \pi^-$   &  $92.4\pm 0.1 $    &$91.7\pm 0.5$ \\
$K^+ K^-$     &  $94.7\pm 0.1 $    &$94.7\pm 0.5$ \\
\hline\hline
\end{tabular}
\end{center}
\caption{Percentage of events with $\left |\de \right |<150\mev$ and photon energy below the cut-off ($2.6 \mev$) 
in the Monte Carlo simulation, as given 
by the (see Tab.~\ref{tab:first_bin}) simulation and by the QED calculation described in the text.}
\label{tab:first_bin}
\end{table}

As explained in Ref.~\cite{QED}, while taking into 
account radiative corrections, one needs to be careful to quote the results in 
such a way that the radiation effects can be disentangled. In principle,
it would be necessary to select $B$ candidates with a specified maximum amount
of ${\cal O}(100\mev)$ photon energy in the final state, a quantity that is
difficult to reconstruct with the \babar\ detector.
Instead, we define our data sample by
selecting on $\de$, an observable that can be  related 
to the maximum allowed total energy of the 
photons, $E_\gamma^{\rm max}$.
The chosen $\de$ window allows for the presence of radiated photons with total energy 
up to $150\mev + \left<\de\right>$, where
the average value of $\de$, $ \left<\de\right>$, differs for each mode, due to the pion mass hypothesis
being assigned to all tracks.  As the $\pip\pim$ events are centered at $\de\sim 0\mev$, 
while the $\Kp \pim$ and $\Kp\Km$ distributions are shifted by $-45\mev$ and $-91\mev$, 
respectively, the corresponding energy requirements on the radiated photons are 
$E_{\gamma}^{\rm max}= 150$, $105$ and $59\mev$ for $\pip\pim$, $\Kp\pim$, and $\Kp\Km$, respectively.
The smearing of $\de$ due to finite momentum resolution leads to a small difference between the 
number of events that satisfy the $\de$ requirement and the number of events that
satisfy an equivalent $E_\gamma^{\rm max}$ requirement. 
We use the MC simulation to evaluate the associated systematic error on the branching fractions from this difference.
%

In addition to signal $\pip\pim$, $\Kp\pim$, and (possibly) $\Kp\Km$ events, the selected data
sample includes background from the process $\epem\to q\bar{q}~(q = u,d,s,c)$.  According
to the MC simulation, backgrounds
from other $B$ decays are small relative to the signal yields ($<1\%$), and are treated as a 
systematic uncertainty.  We use an unbinned, extended maximum-likelihood (ML) fit to extract 
simultaneously signal and background yields in the three topologies ($\pi\pi$, $K\pi$, and $KK$).  
The fit uses the discriminating variables $\mes$, $\de$, the Cherenkov angles of the two tracks, 
and a Fisher discriminant $\cal F$, based on
the momentum flow relative to the $h^+h^{\prime -}$ thrust axis of all tracks and clusters
in the event, excluding the $h^+h^{\prime -}$ pair, as described in Ref.~\cite{BaBarsin2alpha2002}. 
The likelihood for event $j$ is 
obtained by summing the product of the event yield $N_i$ and probability ${\cal P}_i$ over the 
signal and background hypotheses $i$.  The total likelihood for a sample of $N$ events is
\begin{equation}
{\cal L} = \frac{1}{N!}\exp{\left(-\sum_{i}N_i\right)}
\prod_{j}\left[\sum_{i}N_i{\cal P}_{i}(\vec{x}_j;\vec{\alpha}_i)\right].
\end{equation}
The probabilities ${\cal P}_i$ are evaluated as the product of 
the probability density functions (PDFs) with parameters $\vec{\alpha_i}$,
for each of the independent variables 
$\vec{x}_j = \left\{\mes, \de, {\cal F}, \theta_c^+,\theta_c^-\right\}$,
where $\theta_c^+$ and $\theta_c^-$ are the Cherenkov angles for the 
positively- and negatively-charged tracks, respectively.  We check that the variables
are almost independent.  The largest correlation
between the $\vec{x}_j$ is $13\%$ for the pair $(\mes,\de)$, and we have confirmed
that it has a negligible effect on the fitted yields.  For both signal and background, 
the ${\Kpm\pimp}$ yields are parameterized as 
$N_{\Kpm\pimp} = N_{K\pi}\left(1\mp\akpi\right)/2$, and we fit directly
for the total yield $N_{K\pi}$ and the asymmetry $\akpi$.  The result for
$\akpi$ is used only as a consistency check and does not supersede our previously
published result~\cite{babarAkpi}.

The eight parameters describing the background shapes for $\mes$, $\de$, and ${\cal F}$ are
allowed to vary freely in the ML fit.  We use a threshold function~\cite{argus}
for $\mes$ (one parameter),  a second-order polynomial for $\de$ (two parameters), and
a sum of two Gaussian distributions for ${\cal F}$ (five parameters).  For the signal shape in 
$\mes$, we use a single Gaussian distribution to describe all three channels and allow the mean 
and width to vary in the fit.  For $\de$, we use the sum of two Gaussian distributions
(core $+$ tail), where the core parameters are common to all channels and are allowed to vary freely, 
and the tail parameters are determined separately for each channel from the reweighted MC simulation 
(explained above), and fixed in the fit.  
For the signal shape in ${\cal F}$, we use an asymmetric Gaussian function with 
different widths below and above the mean.  All three parameters are determined from MC 
simulation and fixed in the maximum-likelihood fit.  The $\theta_c$ PDFs are obtained from a 
sample of approximately $430000$ $D^{*+}\to D^0\pi^+\,(\Dz\to\Km\pip)$ decays reconstructed in 
data, where $\Km/\pip$ tracks are identified through the charge correlation with the $\pip$ 
from the $D^{*+}$ decay.  We construct the PDFs separately for $\Kp$, $\Km$, $\pip$, and 
$\pim$ tracks as a function of momentum and polar angle using the measured and expected values of 
$\theta_c$, and its uncertainty.  We use the same PDFs for tracks in signal and background events.

Table~\ref{tab:results} summarizes the fitted signal and background yields, and $K\pi$ 
charge asymmetries.  We find a value of $\akpi$ consistent with our previously 
published result~\cite{babarAkpi}, 
and a background asymmetry consistent with zero.  The signal yields are slightly higher than the 
values reported in Ref.~\cite{babarAkpi} due to the removal of the $\Delta t$ selection 
requirement and the addition of the radiative tail in the signal $\de$ PDF.  In order to quantify 
the effect of FSR on the fitted yields, we perform a second fit using a single Gaussian for the 
$\de$ PDF, allowing the mean and width to vary.  The results are shown in the second column 
of Table~\ref{tab:results}, where we find that ignoring FSR lowers the $\pi\pi$ yield by $3.4\%$ 
and the $K\pi$ yield by $1.3\%$.
\begin{table}[!tbp]
\caption{Summary of results from the ML fit for the yields.
The subscript $b$ refers to background.  For the nominal fit, we use a double Gaussian
for the signal $\de$ PDF, as described in the text.  We also show, for comparison purposes, 
the results using a single Gaussian, which corresponds
to an analysis that ignores FSR effects.} 
\begin{center}
\begin{tabular}{ccc}
\hline
\hline
 Parameter                 &	Nominal Fit        & Ignoring FSR\\
\hline
$N_{\pi\pi}$               &$      485 \pm 35     $&$   469\pm 34       $ \\
$N_{K\pi}$                 &$   1656 \pm 52     $&$  1634\pm 52       $ \\
${\cal A}_{K\pi}$          &$  -0.136\pm 0.030  $&$  -0.135\pm 0.030  $ \\
$N_{KK}$                   &$     3  \pm 13     $&$	5 \pm 13   $ \\
$N_{b\pi\pi}$              &$   32983\pm 194    $&$   32998\pm 194    $ \\
$N_{bK\pi}$                &$   20778\pm 169    $&$   20801\pm 169    $ \\
${\cal A}_{bK\pi}$         &$   0.002\pm 0.008  $&$   0.002\pm 0.008  $ \\
$N_{bKK}$                  &$   13358\pm 126    $&$   13356\pm 126    $ \\
\hline\hline
\end{tabular}			
\label{tab:results}
\end{center}
\end{table}

\begin{figure*}[!tbp]
\begin{center}
\includegraphics[width=0.29\linewidth]{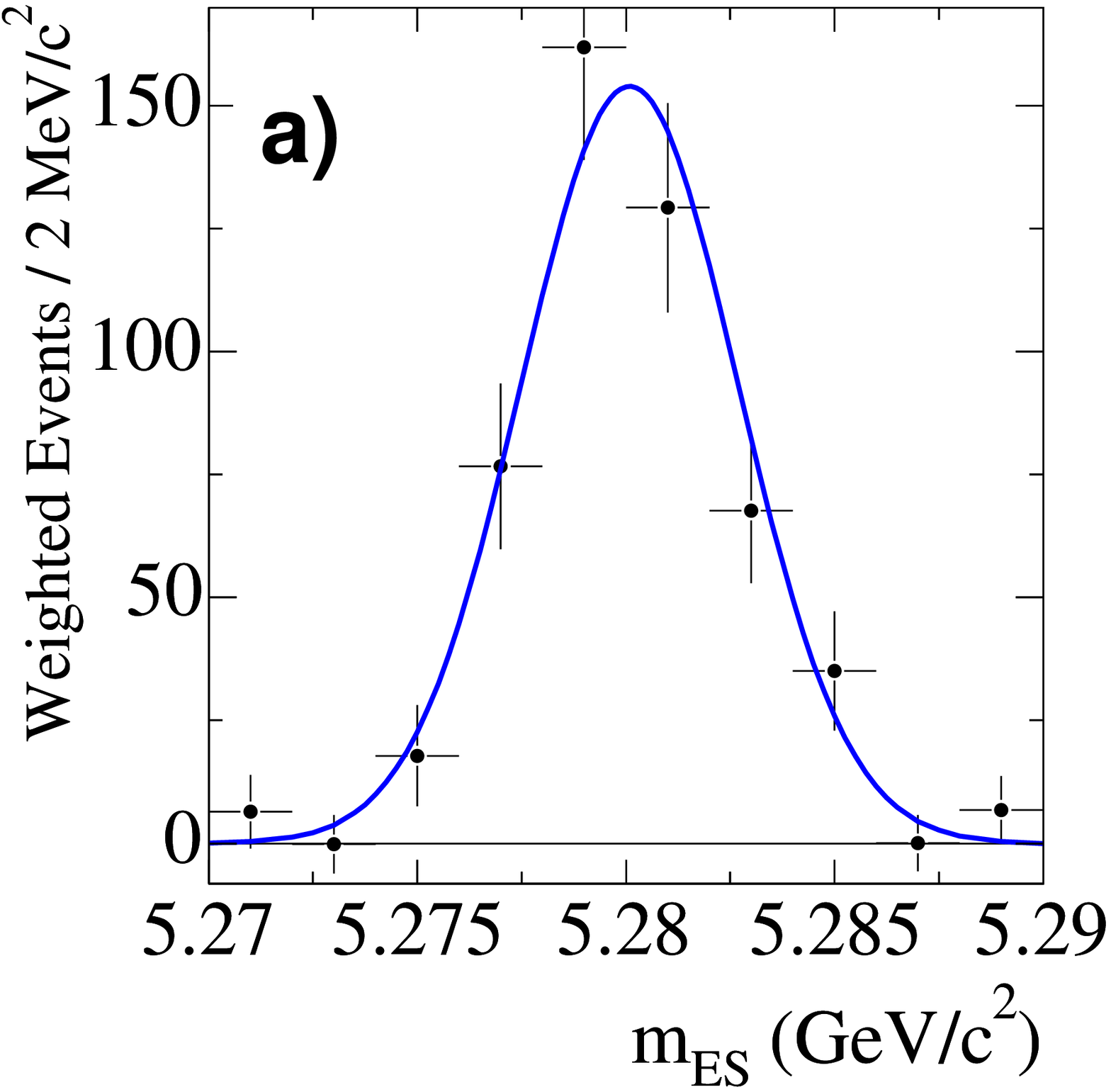}
\includegraphics[width=0.29\linewidth]{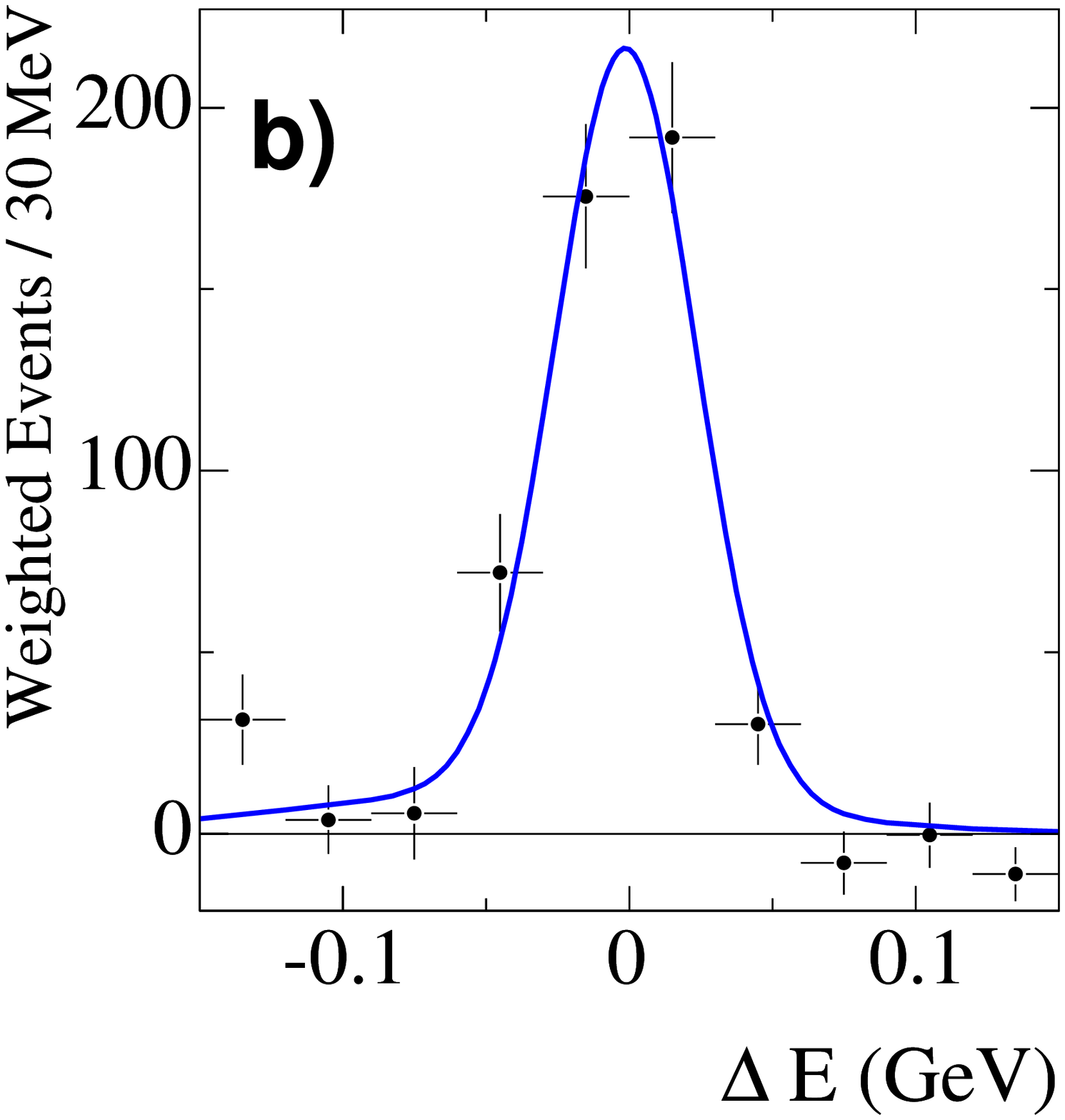}
\includegraphics[width=0.29\linewidth]{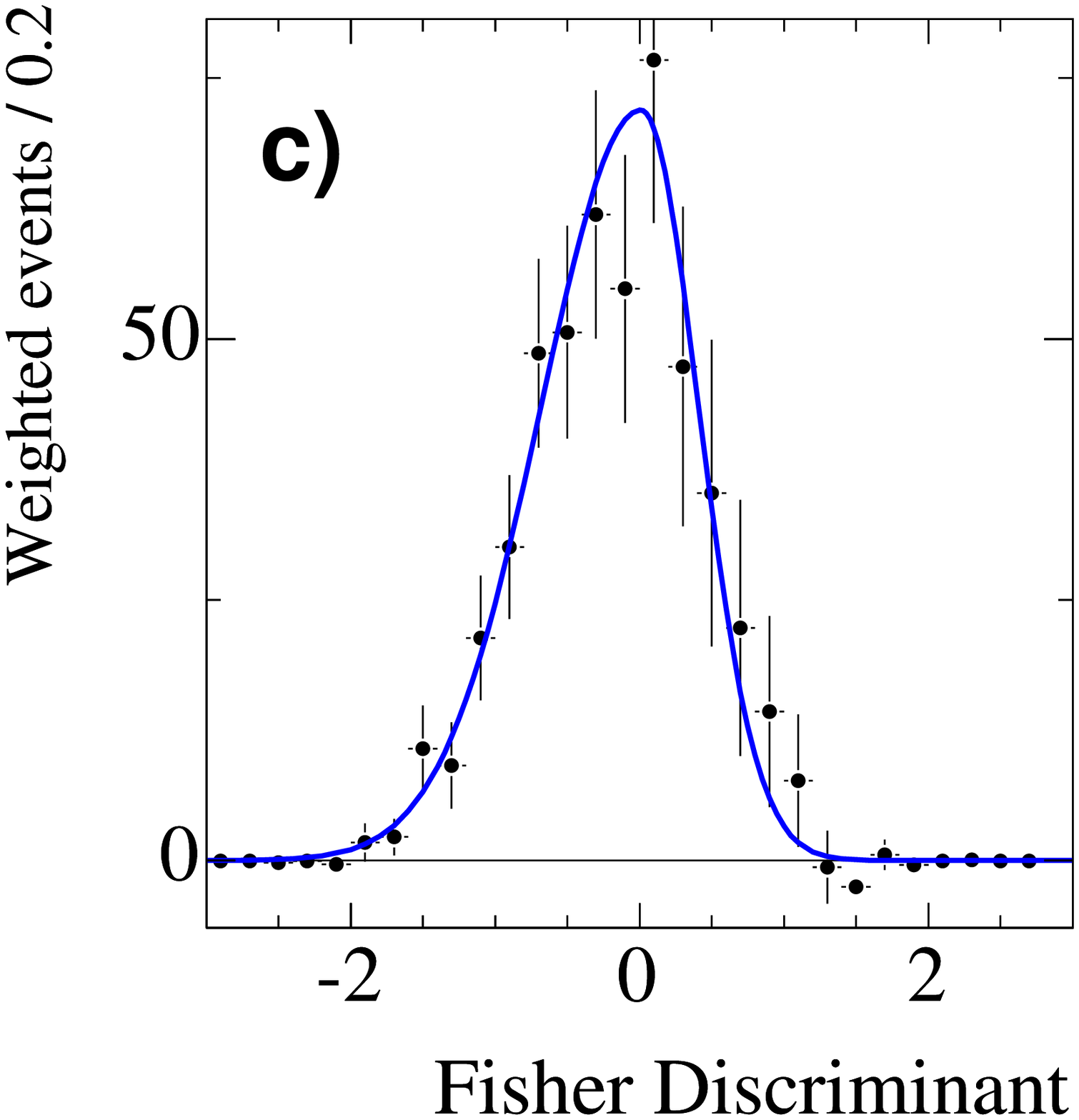}
\includegraphics[width=0.29\linewidth]{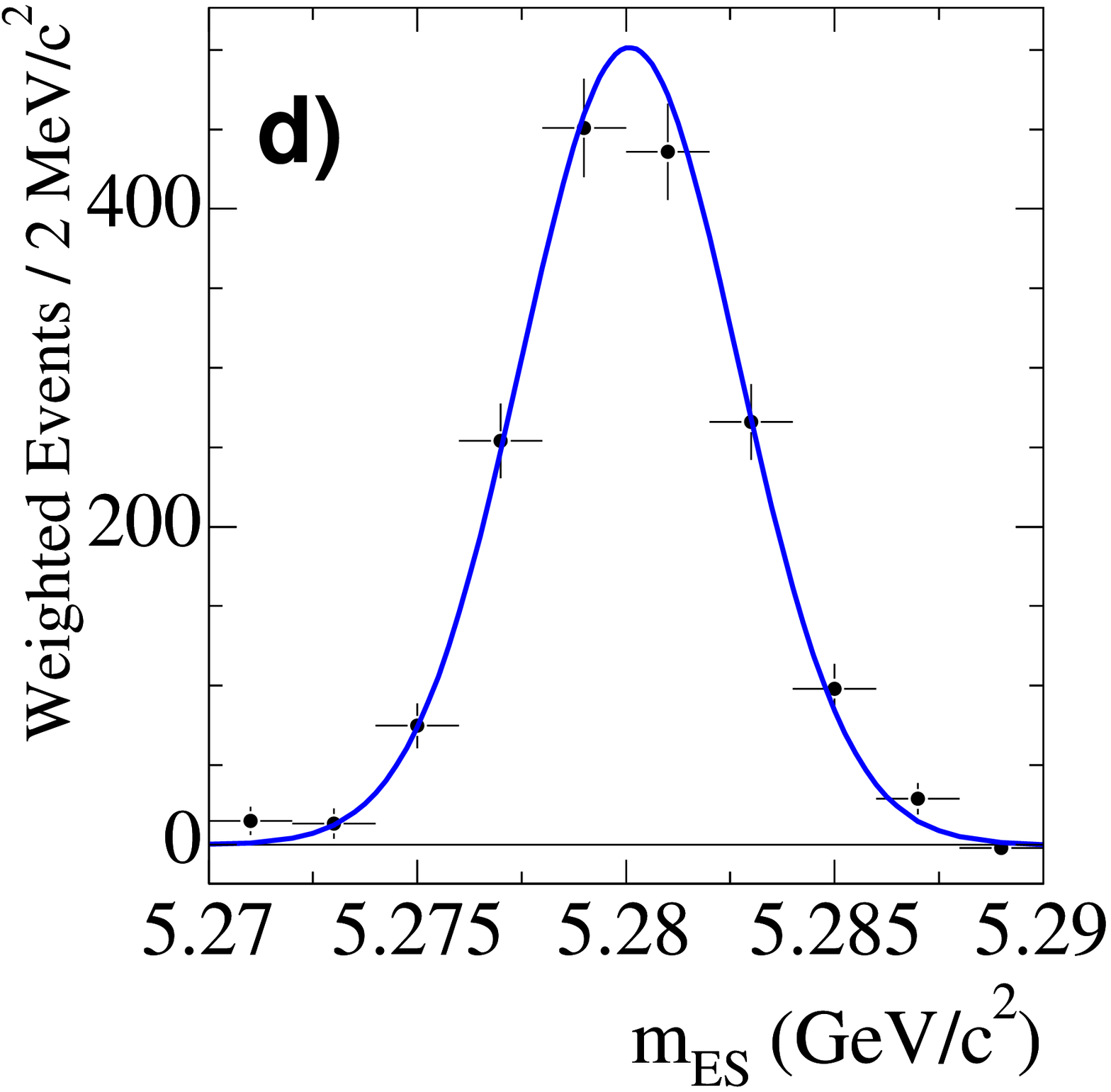}
\includegraphics[width=0.29\linewidth]{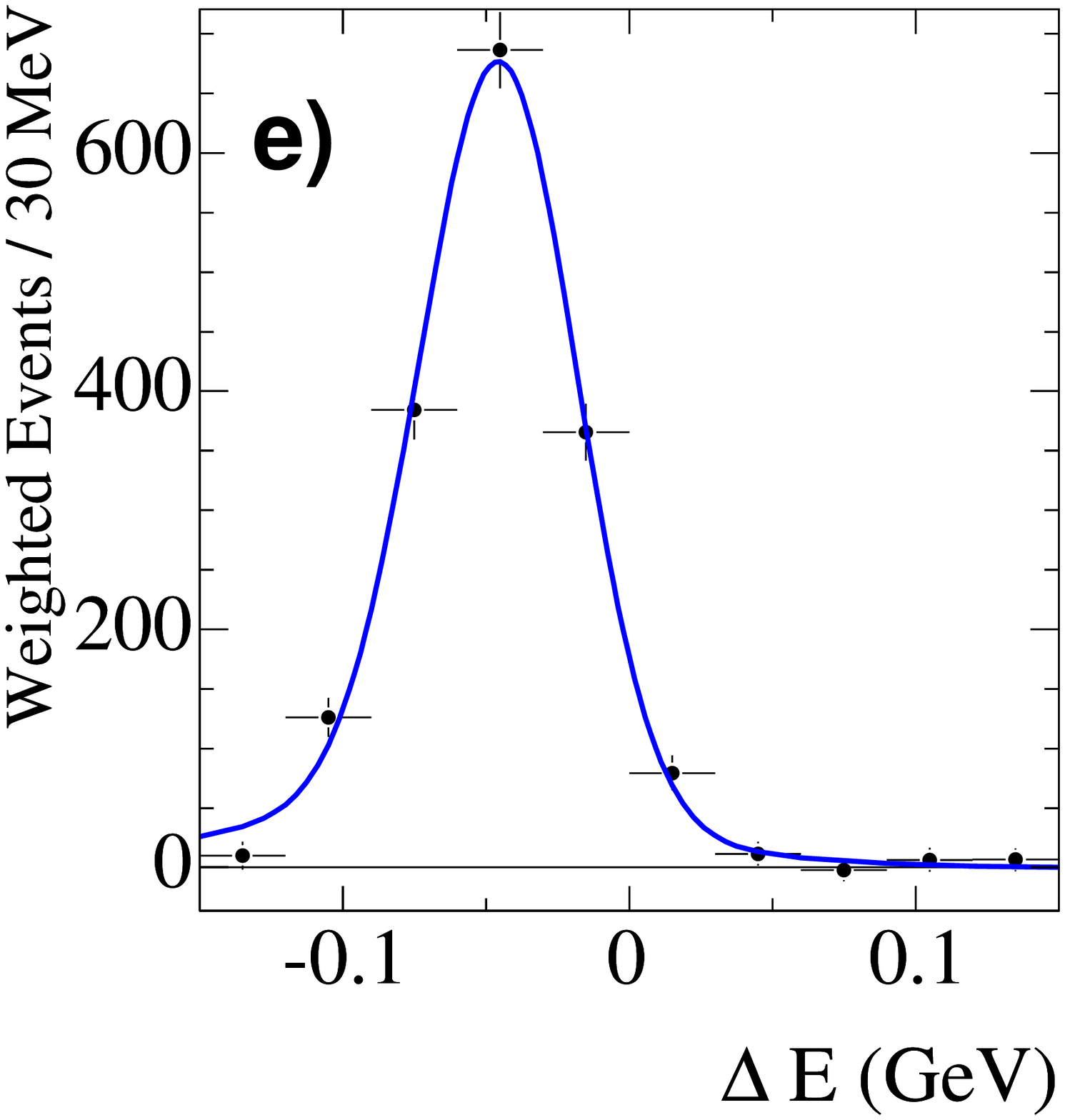}
\includegraphics[width=0.29\linewidth]{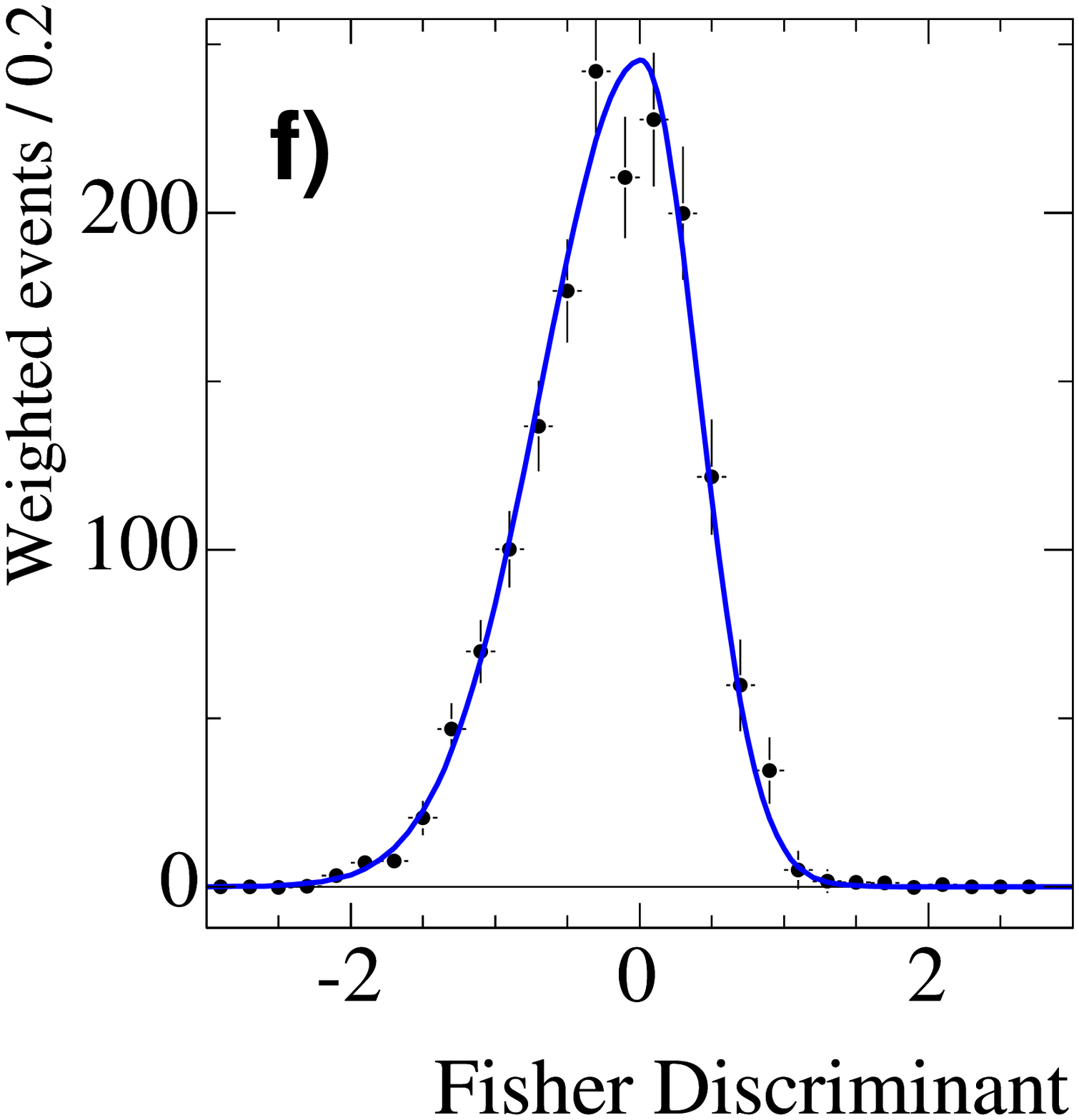}
\includegraphics[width=0.29\linewidth]{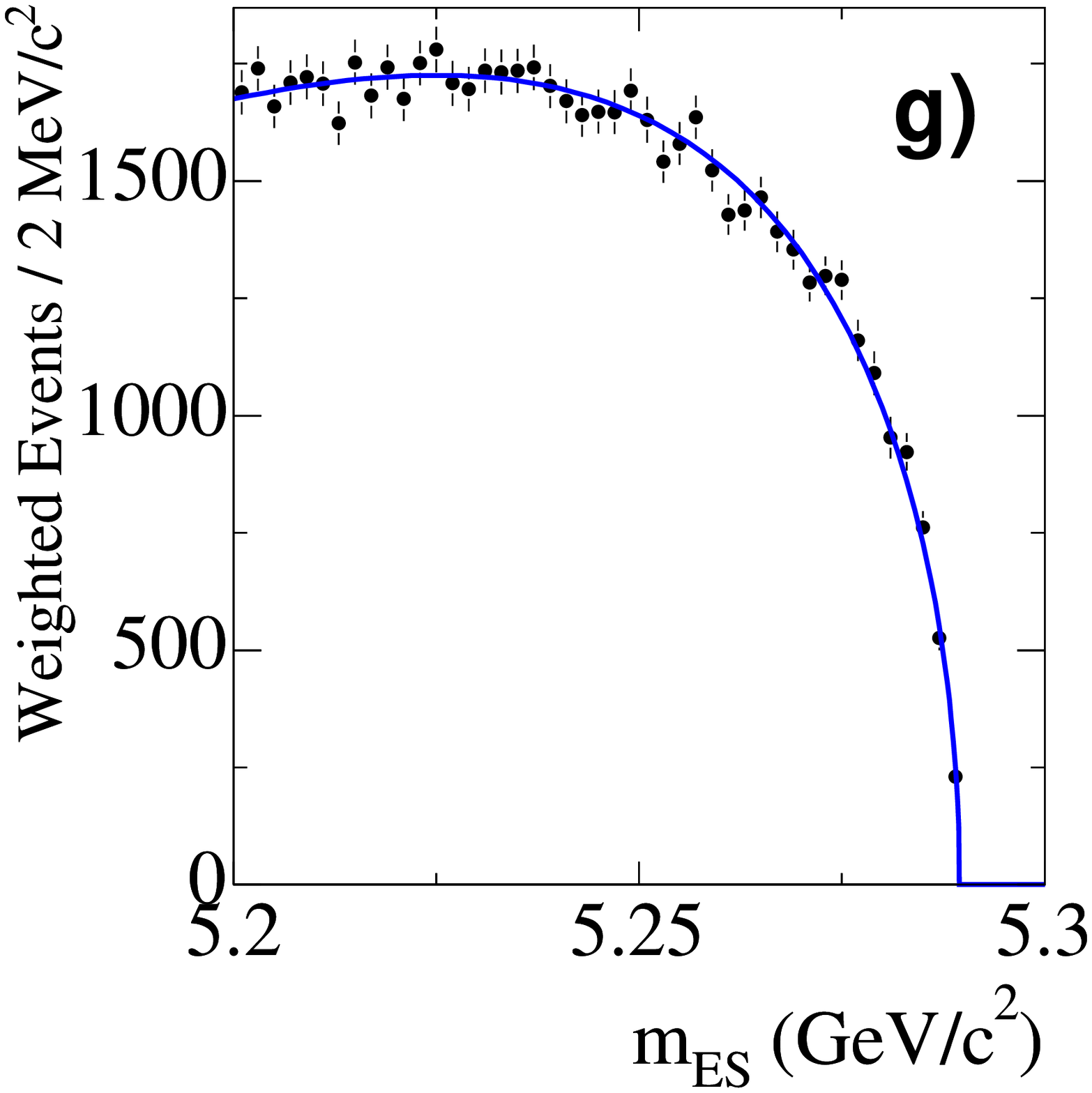}
\includegraphics[width=0.29\linewidth]{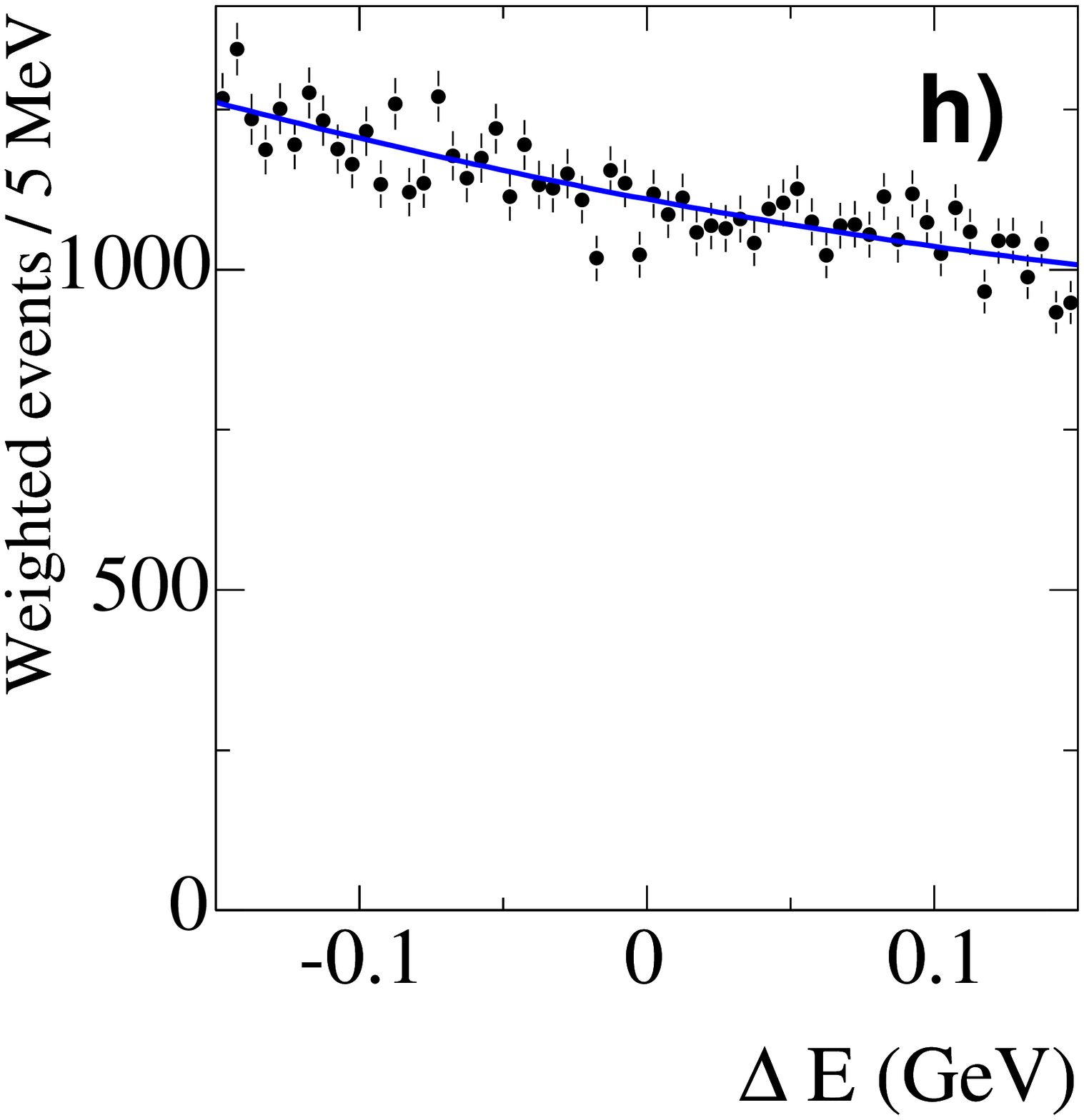}
\includegraphics[width=0.29\linewidth]{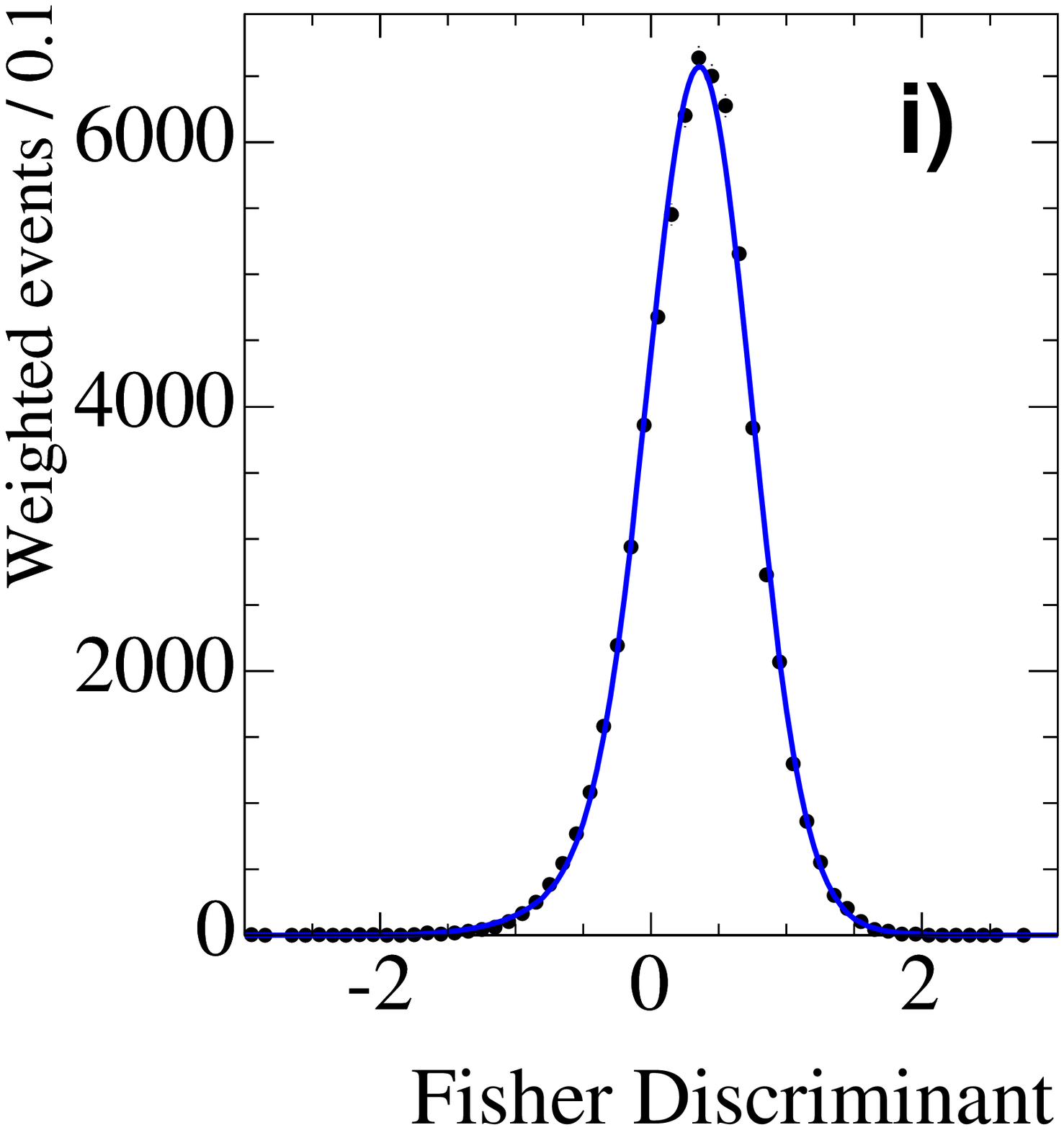}
\caption{Data distributions (points with error bars) of $\mes$, $\de$, and ${\cal F}$ for
signal $\pip\pim$ (a,b,c), signal $\Kp\pim$ (d,e,f) and background for the three
channels (g,h,i), using the weighting technique described in 
the text.  Solid curves represent the corresponding PDFs used in the fit.  The
distribution of $\de$ for signal $\Kp\pim$ events is shifted due to the assignment of the pion mass
for all tracks.}
\label{fig:signal}
\end{center}
\end{figure*}

\begin{figure*}[!tbp]
\begin{center}
\includegraphics[width=0.29\linewidth]{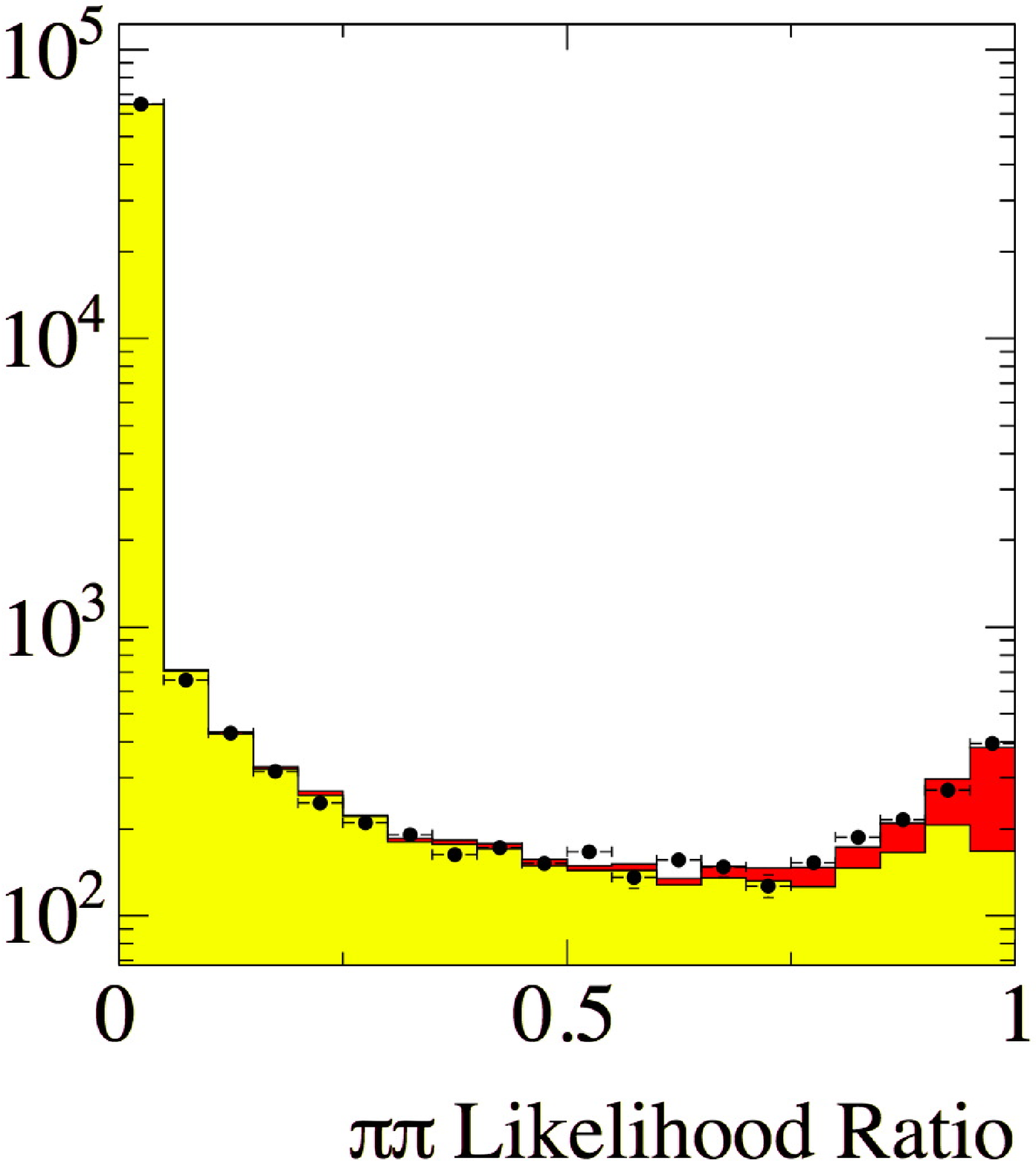}
\includegraphics[width=0.29\linewidth]{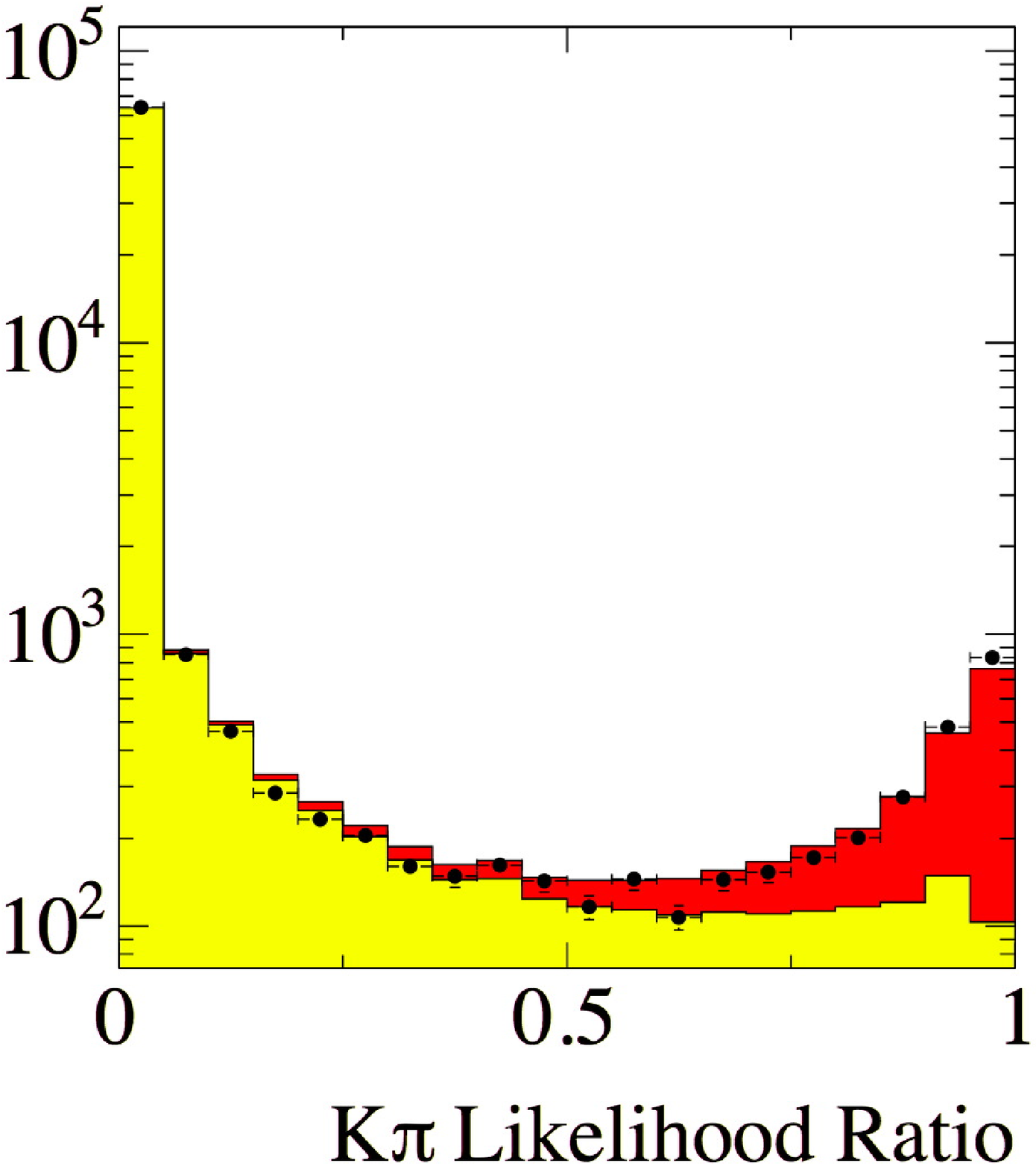}
\includegraphics[width=0.29\linewidth]{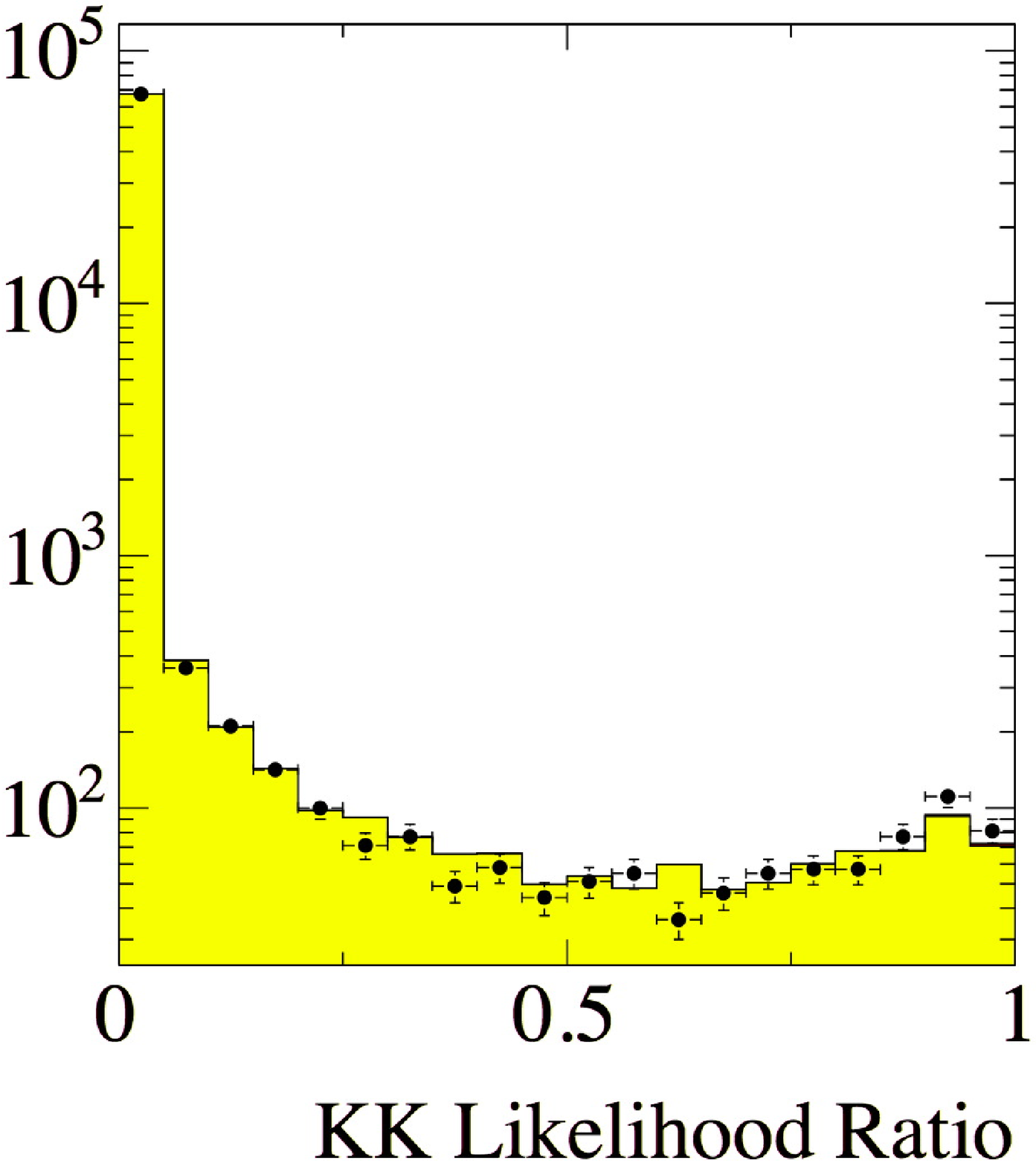}
\caption{(Color online) 
Distribution of the likelihood ratio ${\cal L}_S/\sum{{\cal L}_i}$, where ${\cal L}_S$ is the
likelihood for each event to be a signal $\pip\pim$ (left), $\Kp\pim$ (middle), or $\Kp\Km$ (right) event.
The points with error bars show the distribution obtained on the fitted data sample, while the
histograms show the distributions obtained by generating signal (dark shaded, red) and 
background (light shaded, yellow) events directly from the PDFs.}
\label{fig:lr}
\end{center}
\end{figure*}

As a crosscheck, in Fig.~\ref{fig:signal} we compare the PDF shapes
(solid curves) to the data using the event-weighting technique described in 
Ref.~\cite{sPlots}.  For each plot, we perform a fit excluding the variable being plotted and
use the fitted yields and covariance matrix to determine the relative probability that an event is
signal or background.  The distribution is normalized to the yield for the given component and
can be compared directly to the assumed PDF shape.  We find excellent agreement between
the data and the PDFs.
Figure~\ref{fig:lr} shows the likelihood ratio ${\cal L}_S/\sum{{\cal L}_i}$ for all $69264$ 
events in the fitted sample,
where ${\cal L}_S$ is the likelihood for a given signal hypothesis, and the summation in the
denominator is over all signal and background components in the fit.  We find good 
agreement between data (points with error bars) and the distributions obtained by directly 
generating events from the PDFs (histograms).

Systematic uncertainties on the branching fractions arise from uncertainties on the selection
efficiency, signal yield, and number of $\BB$ events in the sample.  Uncertainty on the 
efficiency is dominated by track reconstruction efficiency   ($1.6\%$) and by the uncertainty on  FSR ($1.3\%$ for $\pi\pi$, $1.4\%$ for $K \pi$ and $2.9\%$ for $KK$), which is evaluated assuming 100\% uncertainty on the smearing effect on $\de$.

Other systematic uncertainties on selection efficiency are those due to requirements on the quality of the $\theta_c$
measurement ($1.0\%$ for $\pi\pi$, $0.8\%$ for $K\pi$ and $0.5\%$ for $KK$) and on event 
topology ($1.1\%$).
Uncertainty on the fitted signal yields is dominated by the shape of the signal PDF for
${\cal F}$ ($2.9\%$ for $\pi\pi$, $1.5\%$ for $K\pi$) and potential bias 
($2.2\%$ for $\pi\pi$, $0.9\%$ for $K\pi$) in the fitting technique, as determined from large 
samples of MC-simulated signal events and a large ensemble of pseudo-experiments 
generated from the PDF 
shapes.  Uncertainties due to imperfect knowledge of the PDF shapes for $\mes$, $\de$, and
$\theta_c$ are all less than $1\%$.  Tables~\ref{tab:sysyield} and \ref{tab:sysbr} summarize the 
uncertainties on the signal yields
and branching fractions, respectively.
\begin{table}[!tbp]
\caption{Summary of relative systematic uncertainties on signal yields.  
For the $\Kp\Km$ yield we show the absolute uncertainty.  
The total uncertainties are calculated as the sum in 
quadrature of the individual contributions.}
\begin{center}
\begin{tabular}{cccc}
\hline\hline
Source           	& $\pip\pim$ $(\%)$ & $\Kp\pim$ $(\%)$ & $\Kp\Km$ (events) \\
\hline
$\mes$                  & $ 0.2$ & $ 0.4$ & $ 1.3 $   \\
$\de$                     & $ 0.1$ & $ 0.0$ & $ 0.3 $ \\
signal $\cal F$         & $ 2.9$ & $ 1.5$ & $ 2.8 $ \\
bkgd $\cal F$           & $ 0.5$ & $ 0.2$ & $ 5.9 $ \\
$\theta_c$ quality      & $ 0.2$ & $ 0.1$ & $ 0.4 $ \\
Fit bias                & $ 2.2$ & $ 0.9$ & $ 1.3 $ \\
$B$ bkgd                & $ 0.8$ & $ 0.2$ & $ < 0.1 $ \\
\hline
Total                   & $ 3.8$ & $ 1.8$ & $6.8$ \\
\hline\hline
\end{tabular}
\label{tab:sysyield}
\end{center}
\end{table}
\begin{table}[!tbp]
\caption{Summary of relative systematic uncertainties on yields, efficiencies, and 
number of $\BB$ pairs.  For the $\Kp\Km$ yield we show the absolute uncertainty.  
The total uncertainties are calculated as the sum in 
quadrature of the individual contributions.}
\begin{center}
\begin{tabular}{cccc}
\hline\hline
Source           	& $\pip\pim$ & $\Kp\pim$ & $\Kp\Km$\\
\hline
yields                  &$ 3.8\% $&$ 1.8\% $&$ 6.8 $ events\\
efficiency              &$ 2.5 \% $&$ 2.5 \% $&$ 3.5 \% $\\
$N_{\BB}$               &$ 1.1\% $&$ 1.1\% $&$ 1.1\% $\\
\hline
Total                   &$ 4.7\% $&$ 3.3\% $& see text \\
\hline\hline
\end{tabular}
\label{tab:sysbr}
\end{center}
\end{table}

Table~\ref{tab:brfinalresults} summarizes the results for the charge-averaged branching fractions.
For comparison, we use the efficiencies and signal yields determined under 
the assumption of no FSR and find $\BR(\Bz\to\pip\pim) = 5.0\times 10^{-6}$ and 
$\BR(\Bz\to\Kp\pim)= 18.0\times 10^{-6}$, which are consistent with our previously published 
results~\cite{BaBarsin2alpha2002}.  
We determine the upper limit for the signal yield for $\Kp\Km$ using a Bayesian procedure that 
assumes a flat prior on the number of events.  The upper limit is given by the value of 
$N_0$ for which 
$\int_0^{N_0} {\cal L}_{\rm max}\,dN/\int_0^\infty {\cal L}_{\rm max}\,dN = 0.90$, 
corresponding to a one-sided $90\%$ confidence interval.  Here, ${\cal L}_{\rm max}$ is the 
likelihood as a function of the $\Kp\Km$ yield $N$, maximized with respect to the remaining fit parameters.  
We find $N_0 = 25.4$, and the upper limit on the branching fraction is calculated by increasing the signal 
yield upper limit and reducing the efficiency by their respective total errors 
(Table~\ref{tab:sysbr}).  For the purpose of combining with measurements by other experiments, 
we also evaluate the central value for the branching fraction and find
$\BR(\Bz\to\Kp\Km n \gamma) = (4\pm 15\pm 8)\times 10^{-8}$.
\begin{table}
\caption{Summary of branching fraction results.  
We give signal yields $N_S$, total detection efficiencies ($\epsilon$) and branching fractions
$\BR_{E_\gamma}$, where the subscript $E_{\gamma}$ serves as a reminder of the dependence on 
the cut on soft photon energy as explained in the text.  
The errors are statistical and systematic, respectively, and
the upper limit on $\Bz\to\Kp\Km n \gamma$ corresponds to the $90\%$ confidence level.}
\begin{center}
\begin{tabular}{lcccc} 
\hline\hline
   Mode    &  $N_S$                  & $\epsilon\,(\%)$ & $\BR_{E_{\gamma}}(10^{-6})$ \\ 
\hline
$\pip\pim$ &  $485\pm 35\pm 18$        & $41.8\pm 0.2\pm 1.0$  & $5.1\pm 0.4\pm 0.2$ \\
$\Kp\pim$  &  $1656\pm 52\pm 30$     & $40.5\pm 0.2\pm 1.0$  & $18.1\pm 0.6\pm 0.6$ \\
$\Kp \Km$  &  $3.2\pm 12.9\pm  7  $    & $39.0\pm 0.3 \pm 1.4$  & $<0.5$ ($90\%$ C.L.) \\
\hline\hline
\end{tabular}
\end{center}
\label{tab:brfinalresults}
\end{table}

Although we cannot directly measure the non-radiative, or ``bare'' branching fractions, due to the 
intrinsic and unavoidable features of QED, they can be extrapolated from our measurements by
employing theoretical calculations, such as those found in Ref.~\cite{QED}. 
The results for these bare branching fractions for the three
channels are shown in Table~\ref{tab:BR_bare}, and the central value
for the bare $K^+ K^-$ branching fraction is
$\BR^0(\Bz\to\Kp\Km) = (4\pm 15\pm 8)\times 10^{-8}$.  We stress the 
importance of being able to disentangle radiation effects from the 
experimental measurements, as a meaningful comparison between theory and experiment 
can be performed only in terms of the bare quantities.  Likewise, bare quantities
should be used when combining measurements from different experiments.
\begin{table}
\caption{Summary of experimental branching fractions, $\BR_{E_\gamma}$, 
with a defined cut on soft photon energy, 
together with the electromagnetic correction factor $G(E_{\gamma}^{\rm max})$ and the 
evaluated ``bare'' branching fractions (non radiative), $\BR^0$.
The errors on branching fractions are statistical and systematic respectively; 
the error on $G(E_{\gamma}^{\rm max})$ is taken as the difference between its 
value at $\mu= M_{\pi}$ and  $\mu= M_{\rho}$.}
\begin{center}
\begin{tabular}{lcccc} 
\hline\hline
   Mode    &   $\BR_{E_{\gamma}}(10^{-6})$& $G(E_{\gamma}^{\rm max})$ & $\BR^0(10^{-6})$ \\ 
\hline
$\pip\pim$ &  $5.1\pm 0.4\pm 0.2$          & $0.937 \pm 0.005$  & $5.5\pm 0.4\pm 0.3$ \\
$\Kp\pim$  &  $18.1\pm 0.6\pm 0.6$       & $0.947 \pm 0.005$  & $19.1\pm 0.6\pm 0.6$ \\
$\Kp \Km$  &  $<0.5$ ($90\%$ C.L.)       & $0.952 \pm 0.005$  & $<0.5$ ($90\%$ C.L.) \\
\hline\hline
\end{tabular}
\end{center}
\label{tab:BR_bare}
\end{table}

In summary, we have presented updated measurements of charge-averaged branching
fractions for the decays $\Bz\to\pip\pim$ and $\Bz\to\Kp\pim$, with FSR effects 
taken into account.  We find that the branching fractions are a few percent higher when the
effect of FSR is included in the calculation of the efficiency and signal yield determination.  
This difference should be taken into account when comparing with previous measurements of these 
quantities~\cite{BaBarsin2alpha2002,belleBR,belleKK,cleoBR} that do not include these effects.  
In order to perform the most meaningful comparison, we also evaluated the
bare branching fractions for the three channels, as explained in Ref.~\cite{QED}.
Our results are consistent with current theoretical estimates from different models~\cite{thy}.  
We find no evidence for the decay $\Bz\to\Kp\Km$ and set an upper limit of $5.0\times 10^{-7}$ at the 
$90\%$ confidence level.

We are grateful for the excellent luminosity and machine conditions
provided by our \pep2\ colleagues, 
and for the substantial dedicated effort from
the computing organizations that support \babar.
The collaborating institutions wish to thank 
SLAC for its support and kind hospitality. 
This work is supported by
DOE
and NSF (USA),
NSERC (Canada),
IHEP (China),
CEA and
CNRS-IN2P3
(France),
BMBF and DFG
(Germany),
INFN (Italy),
FOM (The Netherlands),
NFR (Norway),
MIST (Russia),
MEC (Spain), and
PPARC (United Kingdom). 
Individuals have received support from the
Marie Curie EIF (European Union) and
the A.~P.~Sloan Foundation.


\begin{thebibliography}{99}
\bibitem{cc}
The use of charge conjugate modes is implied throughout this paper unless otherwise
stated.
\end{thebibliography}
\end{document}